# Retail-GPT : leveraging Retrieval Augmented Generation (RAG) for building E-commerce Chat Assistants

**Authors:**
**Bruno Amaral Teixeira de Freitas, IC/Unicamp**
**Prof. Dr. Roberto de Alencar Lotufo, FEEC/Unicamp**

## Introduction:

Large Language Models (LLMs), especially following the release of OpenAI's GPT series, have significantly disrupted textual human-machine interaction. They have enabled the development of chat assistants—also known as chatbots—that engage in more natural conversations and better understand users' needs. When combined with Retrieval-Augmented Generation (RAG) [1] techniques, these models can interact with other software systems and expand the information encoded in their parameters with data retrieved from external sources.

Some examples described in the literature include FACTS [2], NVIDIA's framework for building assistants that leverage enterprise data for enhancing employee productivity and Abbasian et al. [3] health agents, focused on assisting users with healthcare-related tasks.

Another possible domain for such systems is online shopping and delivery services. With estimated global retail e-commerce sales surpassing 6.3 trillion US dollars in 2024 [4], developing alternatives to enhance the customer experience in online purchases holds significant commercial value. In this context, this work describes Retail-GPT, an original open-source RAG-based chatbot designed to guide users through product recommendations and assist with cart operations, aiming to enhance user engagement with retail e-commerce and serve as a virtual sales agent. The goal of this system is to test the viability of such an assistant and provide an adaptable approach for implementing sales chatbots across different retail businesses.

## Methods:

To develop Retail-GPT, the focus was on creating a cross-platform chatbot that could be easily adapted to most e-commerce domains, avoiding reliance on the specific features of any chat application or the particularities of any commercial activity. The objective was to develop a system capable of engaging in natural conversations, mimicking human interaction, and guiding the user through the entire process of making a sale. It was crucial for the chatbot to interpret users' demands accurately, check the availability of products by querying an external product search mechanism, and perform tasks like adding or removing products from users' carts.

One important aspect considered when designing such a system is that LLMs are unpredictable and prone to hallucination [5]. Because of this, our system does not put such models in charge of generating all the responses, nor does it use them for tasks such as returning the users' cart state, processing payment method decisions, or extracting delivery addresses. Another advantage of not using the LLM for generating all the responses is the computational cost saved by, instead, employing a lighter Natural Language Processing (NLP) approach composed by a transformer-based message classifier and pre-written responses.

From the user's perspective, as shown in **Figure 1**, a conversation with Retail-GPT starts with a set of controlled questions, called the initial form, through which data such as delivery addresses, zip codes or registration information is obtained. After this initial step, the chatbot continues engaging in the conversation, providing product recommendations according to the user's demands and performing cart additions or removals according to the user's needs, all independently of buttons or any other form of restraining the format of the users' messages. Finally, when the user indicates that their cart is complete, the chatbot starts another set of controlled questions regarding the checkout process of the sale, with the confirmation of the cart and payment method specification. The idea is that a business could customize such forms depending on the information it requires.

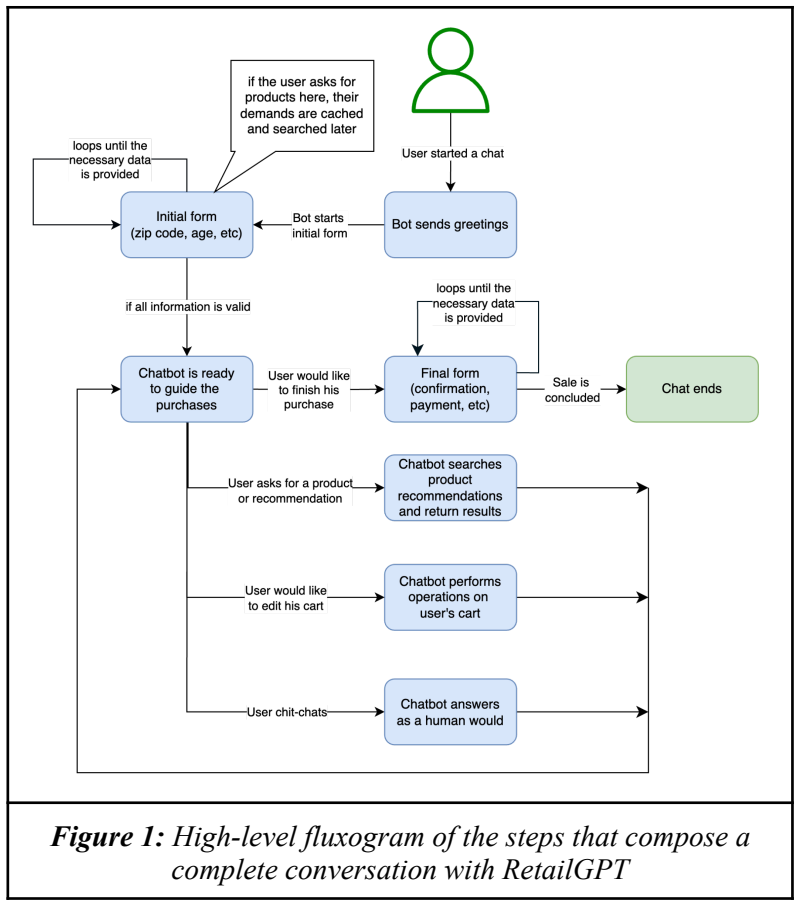

*Figure 1: High-level fluxogram of the steps that compose a complete conversation with RetailGPT*

To achieve this, a combination of RAG and a NLP pipeline built around a message classifier was implemented. As shown in **Figure 2**, the user message is first tokenized and then converted into vector embeddings by the featurizer. These embeddings are fed into a DIET (Dual Intent and Entity Transformer) classifier [6], which extracts data (entities) and determines the response—either sampling a pre-written message or generating one with the LLM-based subsystem. The chatbot uses a sequence of pre-established questions to gather necessary data, with the transformer model extracting information and deciding how to respond based on user interaction. While effective, this NLP pipeline alone can't handle the vast possibilities of user messages or produce contextually relevant, human-like responses.

Therefore, the classifier can delegate message processing to the LLM-based subsystem when needed. The message first passes through input guardrails to block inappropriate content or sensitive data using moderation models and an LLM designed to detect jailbreak attempts. If secure, the message is added to the system's main prompt, which includes general instructions and defines function calls for product search, cart editing, and purchase completion, as well as few-shot examples [7] demonstrating expected behaviors. Previous interactions stored in a database are also appended to provide memory. The LLM's textual response is then checked for appropriateness before being returned to the user. Alternatively, the LLM generates one of three function calls: purchase completion, cart operations, or product search.

Purchase completion triggers the chatbot's final form, cart operations invoke handlers for adding or removing items from the cart, and product search calls pass a query to a search mechanism that generates recommendations. This search mechanism is mocked by calling an LLM with a fictional dataset. The chatbot can process multiple function calls concurrently for efficiency. The cart operation handler verifies product availability in the database, searches for it if necessary, and requests user confirmation before any addition. A cart summary, generated without the LLM to avoid errors, is sent after modifications.

The entirety of the system was built with Python 3. The NLP pipeline was implemented through Rasa, an open-source framework for building chatbots with machine-learning, using SpaCy's, an open-source library of NLP, en_core_web_lg pipeline. The chosen LLM was GPT-4o, accessed via OpenAI's API. For demonstration purposes, the system was developed as an assistant for a fictional Foo convenience store, featuring around 50 products related to the retail segment analyzed, that were commonly found in their stores. The initial form was configured to only ask for basic customer's identification information. The final form was configured to only ask for a confirmation of the purchase and the user's preferred payment method. For the database, Redis was used. Finally, the search mechanism was mocked by simply calling the LLM with a prompt asking for a recommendation based on the available products.

### Results and Discussion:

The source code can be accessed through https://github.com/unicamp-dl/retailGPT. During qualitative testing, it was verified that RetailGPT is capable of engaging in conversations that fulfill the objectives described previously in the Methods section. **Figures 3 and 4** show real conversations that exemplify the system's capabilities in guiding a sale through cart operations and product recommendations.

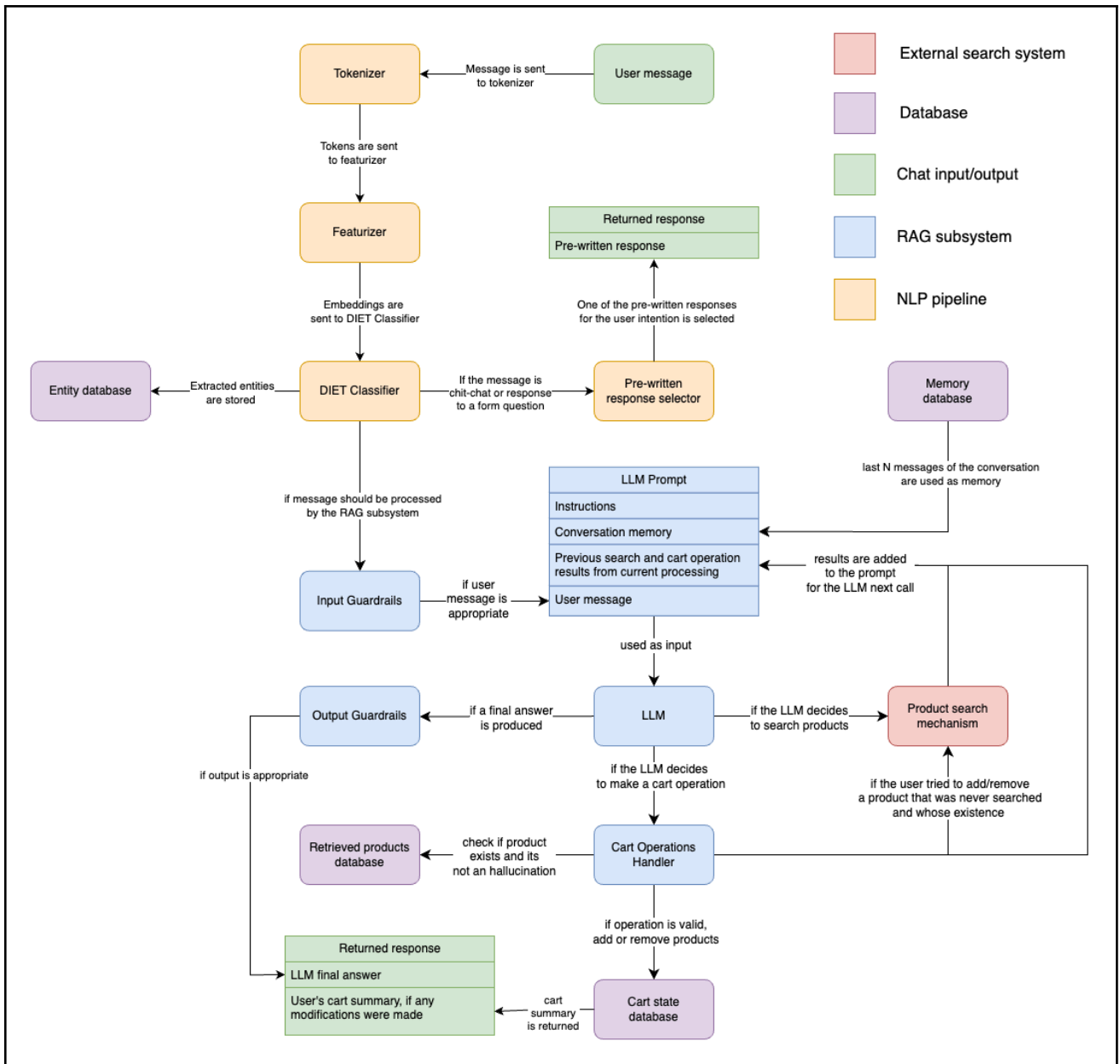

*Figure 2: Retail-GPT message processing fluxogram*

Despite the promising results, a non negligible amount of user's messages during testing resulted in errors by the LLM, that was not capable of generating the appropriate function call or interpreting the subtleties of the user's intentions, making RetailGPT still not reliably suitable for production environments. Alternative approaches for prompting the model could be tested in future works in order to minimize this problem. In this sense, adding a Chain-of-Thought (CoT) [8] reasoning step before the production of the final answer or the function calls could enhance the system robustness. Besides that, ReAct [9] is a RAG technique that could substitute the function calling approach, at the same time it adds a reasoning step. However, such modifications could, at best, minimize the LLM's errors, with no guarantee of eliminating them.

Additionally, the system also has limitations regarding its ability to deal with unexpected scenarios, such as if the user asks the LLM to change his zip code, directly return the state of the cart or provide more in depth information about any of the products. More functions could be added to the prompt to encompass such interactions, at the cost of increasing the complexity of the task delegated to the LLM.

While it is difficult to quantitatively evaluate an application like RetailGPT—due to the impracticality of testing all the infinite ways a user could interact with the system and the heavy reliance on external software like the chosen LLM and search mechanism (mocked in our implementation)—the experiments below were conducted to test the RAG subsystem, which has been identified as the main source of the issues mentioned earlier. In all experiments, the chatbot started with an empty memory for each test case.

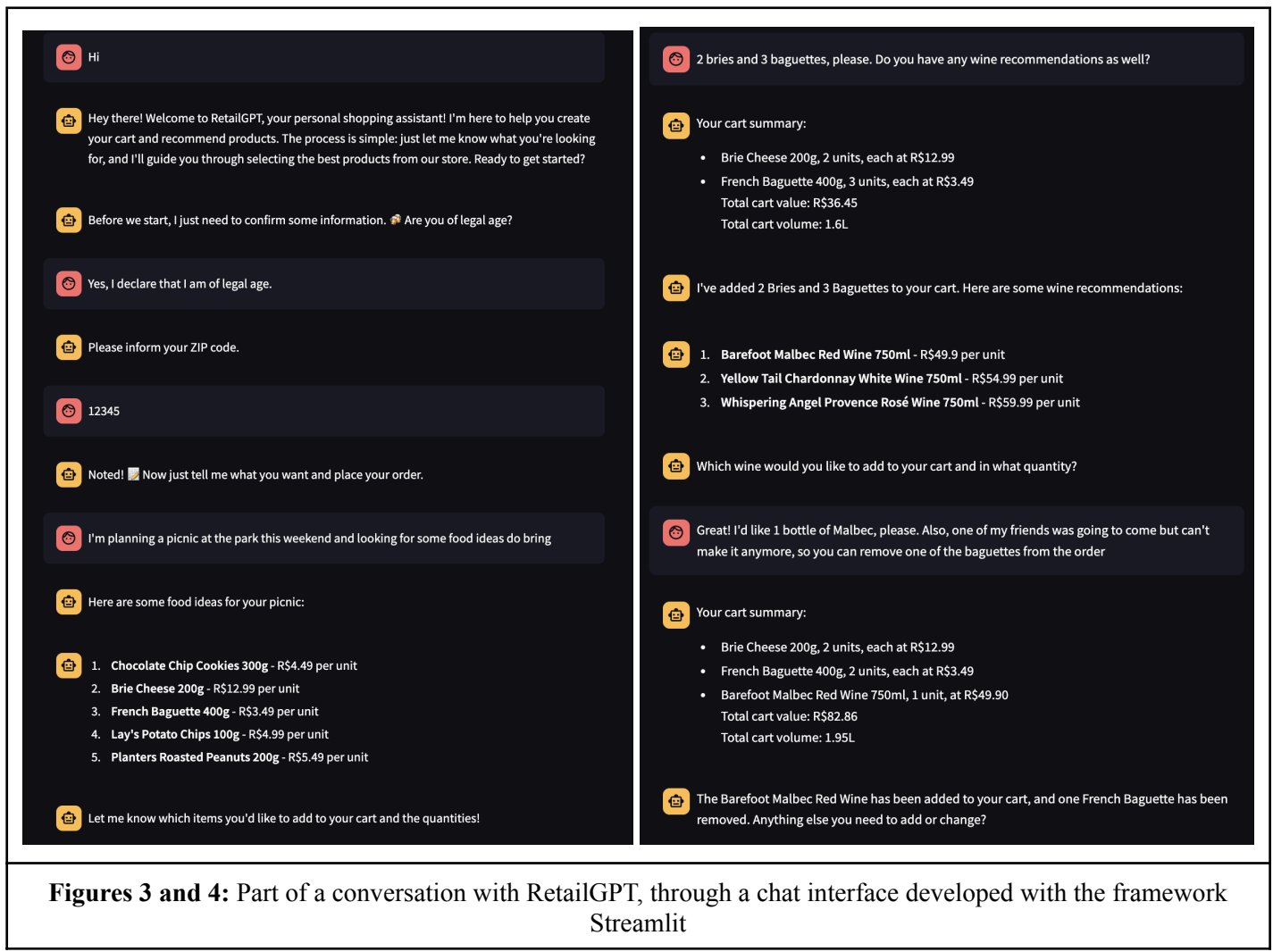

**Figures 3 and 4:** Part of a conversation with RetailGPT, through a chat interface developed with the framework Streamlit

Firstly, the system's ability to correctly select the appropriate tool—whether cart editing (adding or removing products), product search, finalizing a purchase, or responding without any tool—was tested using a handmade dataset. This dataset matched fictitious scenarios and user messages to the corresponding tools that should be invoked. For each tool usage scenario, 20 tests were conducted, including an additional 20 tests for messages that required both a product search and a cart edit, which is also a common case, as shown in **Figures 3 and 4**. For the cart edit tests, half of the operations involved additions and the other half removals.

As shown in **Table 1**, the chatbot handled messages that triggered product searches or required no tool usage correctly. However, during cart removals, there was an instance where the LLM did not follow the correct structure for the tool's arguments, indicating a vulnerability to hallucinations. The lower accuracy in cart additions was due to unnecessary product searches to confirm product existence, even when the chat history already indicated that the product was present. Although these cases were not highly problematic—since the chatbot eventually added the product after the extra search during the next interaction—they highlight potential unexpected behaviors in LLM-based applications. A similar issue occurred with the finishing purchase completion tool, where the chatbot asked for user confirmation instead of inferring intent and calling for finishing the order directly. Moreover, the results show that the system can handle messages requiring both searches and cart operations with similar accuracy to cases where only one of these operations is performed.

Overall, the results suggest that the LLM can generally invoke the correct tools with the current prompt, with only one significant hallucination out of 80 test instances. However, it's important to note that the dataset consists of idealized messages and scenarios, which may not capture the full complexity of real-world interactions or uncover other potential issues.

| Tool | Accuracy |
|---|---|
| Product search | 100% |
| Cart addition | 60% |
| Cart removal | 90% |
| No action | 100% |
| Finish purchase | 95% |
| Product search + cart addition | 70% |
| Product search + cart removal | 100% |

**Table 1**: Results for tool call tests

| Test | Accuracy |
|---|---|
| Prompt injections | 75% |
| Corner-cases | 100% |
| Off-topic | 100% |

**Table 2:** Results for security and consistency tests

Secondly, the security and consistency of the RAG subsystem were tested using another handmade dataset, which included 20 prompt injections – attempts to insert malicious instructions or information into the prompt, 20 corner-case messages – requesting nonexistent products or actions the chatbot cannot perform – and, finally, 20 messages unrelated to the chatbot's purpose, where the assistant should redirect the user back to making a purchase.

As shown in **Table 2**, the chatbot handled off-topic messages and requests for unavailable information or tools well, consistently redirecting the user to its intended functions. However, these results serve only as an indicator of quality and may not fully reflect real-world performance, as the system could fail when faced with more complex scenarios. Notably, in 25% of the prompt injections, the guardrails system and the chatbot prompt were insufficient to prevent RetailGPT from being misled into believing in nonexistent products or offering unrealistic discounts.

Such results show that, while the system acts consistently in the majority of cases, it is not difficult for a malicious user to hack the system. In this sense, future approaches must develop more rigid guardrails in order to make the system secure enough for production environments.

## Conclusions:

This work described Retail-GPT, an open-source RAG-based approach for building conversational chatbots for retail e-commerce. The system is built around a DIET classifier and a large language model (LLM) connected to an external search engine and databases. Our implementation specifically used the Python framework Rasa, Redis, SpaCy's en_core_web_lg pipeline, and OpenAI's GPT-4o, both as the system's LLM and to simulate the external search mechanism.

Our results indicate that while Retail-GPT can manage conversations, guide users through purchases by extracting personal information, make product recommendations, and perform cart operations, it is not free from hallucinations or security issues. The approach demonstrates that although RAG has the potential to disrupt retail, there are still challenges to be addressed in future works in order to build an application suitable for production environments.

## References:

[1] LEWIS, Patrick et al. **Retrieval-Augmented Generation for Knowledge-Intensive NLP Tasks**. arXiv, 2020.
[2] AKKIRAJU, Rama et al. **FACTS About Building Retrieval Augmented Generation-based Chatbots**. arXiv, 2024
[3] MAHYAR, Abbasian et al. **Conversational Health Agents: A Personalized LLM-Powered Agent Framework**. arXiv, 2024
[4] STATISTA. **E-commerce worldwide - statistics & facts**. Available at: https://www.statista.com/topics/871/online-shopping/#topicOverview. Accessed on: 28/07/2024.
[5] XU, Ziwei; JAIN, Sanjay; KANKANHALLI, Mohan. **Hallucination is Inevitable: An Innate Limitation of Large Language Models**. arXiv, 2024.
[6] BUNK, Tanja; VARSHNEYA, Daksh; VLASOV, Vladimir et al. **DIET: Lightweight Language Understanding for Dialogue Systems**. arXiv, 2020.
[7] BROWN, Tom B.; MANN, Benjamin; RYDER, Nick et al. **Language Models are Few-Shot Learners**. arXiv, 2020.
[8] WEI, Jason; WANG, Xuezhi; SCHUURMANS, Dale et al. **Chain-of-Thought Prompting Elicits Reasoning in Large Language Models**. arXiv, 2022.
[9] YAO, Shunyu; ZHAO, Jeffrey; YU, Dian et al. **ReAct: Synergizing Reasoning and Acting in Language Models**. arXiv, 2022